\documentclass{PoS}

\usepackage{amsfonts}
\usepackage{amssymb}
\usepackage{amsmath}
\usepackage{amsthm}
\def\one{\mbox{1 \kern-.59em {\rm l}}}
\usepackage{cite}

\title{Gravity as a Gauge Theory on Three-Dimensional Noncommutative spaces\footnote{Based on the proceeding for a talk given in 9/17 at the COST Training school ``Quantum spacetime and physics models'' in Corfu, Greece}}

\ShortTitle{3-d Noncommutative Gravity}

\author{D. Jurman$^a$, \speaker{G. Manolakos$^b$}, P. Manousselis$^b$, G. Zoupanos$^{bc}$ \\ \llap{$^a$} Theoretical Physics Division, Rudjer Bo$\check s$kovi\'c Institute, Bijeni$\check c$ka 54, 10000  Zagreb, Croatia\\ \llap{$^b$}Physics Dept, National Technical University, Zografou Campus, GR-15780 Athens, Greece\\ \llap{$^c$}Max-Planck Institut f\"ur Physik, F\"ohringer Ring 6, D-80805 Munich, Germany \\ E-mail: \email{Danijel.Jurman@irb.hr}, \email{gmanol@central.ntua.gr}, \email{pantelis.manousselis@gmail.com}, \email{zoupanos@mail.cern.ch}}

\abstract{We plan to translate the successful description of three-dimensional gravity as a gauge theory in the noncommutative framework, making use of the covariant coordinates. We consider two specific three-dimensional fuzzy spaces based on SU(2) and SU(1,1), which carry appropriate symmetry groups. These are the groups we are going to gauge in order to result with the transformations of the gauge fields (dreibein, spin connection and two extra Maxwell fields due to noncommutativity), their corresponding curvatures and eventually determine the action and the equations of motion. Finally, we verify their connection to three-dimensional gravity.}

\FullConference{Corfu Summer Institute 2017 'School and Workshops on Elementary Particle Physics and Gravity'\\
		2-28 September 2017\\
		Corfu, Greece}

\begin{document}

\section{Introduction}

It has been argued that the consideration of commutativity of coordinates at arbitrarily small length scales is not an obvious and natural choice. Therefore, one could consider that, at small distances, coordinates of spacetime exhibit noncommutative behaviour. A direct query that emerges is whether noncommutativity of coordinates induces modifications for gravity. A modest approach is to consider an alternative, noncommutative version of (first-order formulation of) general relativity and find out the modifications or corrections to the well-established results of general relativity. 

The starting point is to recall the relation between gravity and gauge theories \cite{Utiyama:1956sy, Kibble:1961ba, MacDowell:1977jt, Kibble:1985sn}; general relativity, with or without cosmological constant, is obtained upon gauging the Poincar\'e or (A)dS algebra. This is valid for the transformation rules of the gauge fields---the vielbein and the spin connection---in arbitrary dimensions. However, specifically in three dimensions, besides the transformation rules, the dynamics, i.e. the Einstein-Hilbert action, can be obtained by gauge theory \cite{AchucarroTownsend,WittenCS}. Given that noncommutative gauge theories exist and are well-defined \cite{Madore:2000en}, one could then be motivated to make use  of them as a guide to noncommutative gravity. Such an approach was followed before, for example in Refs. \cite{Chamseddine:2000si,Chamseddine:2003we,Aschieri:2009ky,Aschieri2,Ciric:2016isg}. In a similar way, this has been studied also in three dimensions, using the relation to Chern-Simons gauge theory \cite{Cacciatori:2002gq,Cacciatori:2002ib,Aschieri3, Banados:2001xw}. The above works share a common feature, that is the  noncommutative deformation is constant (Moyal-Weyl) and the whole procedure is followed by using the corresponding star product of functions and the Seiberg-Witten map \cite{Seiberg:1999vs}.

Alternatively, there is another type of noncommutative geometries, matrix geometries, which can be employed in order to study quantum gravity \cite{Banks:1996vh,Ishibashi:1996xs}. Several approaches have been suggested, mainly based on Yang-Mills matrix models \cite{Yang:2006dk,Steinacker:2010rh,Kim:2011cr,Nishimura:2012xs,Furuta:2006kk,Hanada:2005vr,Aoki:1998vn,Nair:2006qg,Abe:2002in,Valtancoli:2003ve,Nair:2001kr}, pointing once more at direct relations among noncommutative gauge theories and gravity. For another approach see Refs. \cite{Buric:2006di,Buric:2007zx,Buric:2007hb}, where a solid indication that the degrees of freedom of the resulting theory of gravity can be put in correspondence with those of the noncommutative structure. In this case, the usual symmetries such as coordinate invariance are built-in, and the commutator of coordinates can have arbitrary dependence on them. Moreover, describing gravity as a gauge theory in the context of matrix geometry is further motivated by gauge theories defined on fuzzy spaces \cite{Aschieri:2003vyAschieri:2004vhAschieri:2005wm}. It is worth noting that the dimensional reduction of higher-dimensional gauge theories over fuzzy manifolds -used as extra dimensions- leads to renormalizable four-dimensional theories \cite{Aschieri:2006uwSteinacker:2007ay}. This is a delicate and desirable feature that is worth exploring in the case of gravity, too.

In general, formulating gravity in the noncommutative framework is a rather difficult task, because noncommutative deformations break Lorentz invariance. Nevertheless, there are certain types of noncommutative spaces, on which it is possible to define deformed symmetries which are preserved, as for example in the case of $\kappa$-Minkowski spacetime \cite{Lukierski:1991pn,Lukierski:1992dt}, which appears as a solution of the Lorentzian IIB matrix model in Ref. \cite{Kim:2011ts}. However, there are special types of deformations that constitute covariant noncommutative spacetimes \cite{Snyder:1946qz,Yang:1947ud}. Studies on such spaces can be found in Ref. \cite{Heckman:2014xha}, where the authors build a general conformal field theory defined on covariant noncommutative versions of four-dimensional dS or AdS spacetime. More four-dimensional constructions of the same spirit are studied in Refs. \cite{Buric:2015wta,Sperling:2017dts,Steinacker:2016vgf,Buric:2017yes}.

In this paper we review the study of three-dimensional gravity as a noncommutative gauge theory \cite{Chatzistavrakidis:2018vfi}. First thing to do is the determination of the three-dimensional noncommutative spaces with the appropriate symmetry. We employ two structures for the Euclidean and the Lorentzian signature. In the Euclidean case, such a space is the foliation of three-dimensional Euclidean space by fuzzy spheres \cite{Hammou:2001cc}, called $\mathbb{R}_\lambda^3$. The noncommutative coordinates of this space belong to an SU(2) algebra, but, in contrast to the fuzzy sphere case \cite{Madore:1991bw, Hoppe}, the hermitian matrices are not restricted to live in unitary irreducible representations but in reducible ones. This means that each coordinate of $\mathbb{R}_\lambda^3$ is written as a matrix in a block diagonal form, with each block being an irreducible representation, i.e. a fuzzy sphere. It is in this sense that a third dimension emerges and the noncommutative space can be viewed as a foliation of the three-dimensional Euclidean space, where each fuzzy sphere is considered as a leaf of the foliation. Then, having determined the space, we are going to gauge its group of symmetries, SO(4) (see for example \cite{Kovacik:2013yca}), in order to develop a model for three-dimensional gravity, combining the existing tools of doing gauge theories on noncommutative spaces \cite{Madore:2000en} and the whole procedure for building three-dimensional gravity in the commutative case \cite{WittenCS}. A generic feature of noncommutative non-Abelian gauge theories is that anticommutators do not close, therefore, in order to overpass this drawback, we will consider a larger symmetry, U(2)xU(2) in a fixed representation. 
The above structure has a Lorentzian analogue, in which a three-dimensional noncommutative space whose coordinates belong to reducible representations of SU(1,1) is involved. This space is a similar construction to the $\mathbb{R}_\lambda^3$, i.e. a foliation of the three-dimensional Minkowski spacetime by fuzzy hyperboloids\footnote{The fuzzy hyperboloid space is obtained after restricting the matrices of SU(1,1) live in unitary irreducible representations (in analogy with the SU(2) case, in which fuzzy sphere is obtained in the same way).} \cite{Jurman:2013ota}. Since the coordinates are matrices in reducible representations of SU(1,1), they are written in block diagonal form of irreducible representations, with each block being a fuzzy hyperboloid, which means that the space we end up with is a foliation of the three-dimensional Minkowski spacetime by fuzzy hyperboloids. Again we proceed in a similar way, gauging the symmetry group of the space, which is the SO(1,3). For the same reason as in the Euclidean case, we have to enlarge the symmetry to the GL(2,$\mathbb{C}$) in a fixed representation (as in \cite{Aschieri:2009ky}).   

The outline of the proceeding is the following. In section 2, we briefly review the construction of four-dimensional gravity as a gauge theory of the Poincar\'e algebra and three-dimensional gravity as a gauge theory of the Poincar\'e or (A)dS algebra, depending on the presence and sign of the cosmological constant, since this is the basic prescription we will use later for the noncommutative case. In section 3, we include some worth-noting remarks on noncommutative gauge theories (emphasizing on the non-Abelian case), since those will be the tools we are going to use for our purpose, but also to make the present text self-contained. In section 4, we give some information about the structure of the noncommutative spaces we are going to work on, i.e. the foliated three-dimensional Euclidean and Minkowski spacetimes by fuzzy spheres and fuzzy hyperboloids, respectively. In section 5, we proceed with building the gauge theories on those spaces, obtaining expressions for the transformation rules of the gauge fields and their curvature tensors and taking the commutative limit in order to compare our findings with the well-established results of the commutative three-dimensional theory of gravity. In section 6, we write down an action of Chern - Simons type and vary it in order to obtain the equations of motion. In section 7, we summarize our work, discuss our findings and propose potential directions for future work.     

\section{Gravity as a gauge theory}

In this section we briefly review the alternative description of gravity in four and three dimensions as gauge theories of their corresponding isometry groups.  

\subsection{Four-Dimensional case}
In four dimensions, the vielbein formulation of general relativity can be understood, at least at the kinematical level\footnote{This is also true in arbitrary dimensions.}, as a gauge theory \cite{Utiyama:1956sy, Kibble:1961ba, MacDowell:1977jt, Kibble:1985sn} of the Poincar\'e algebra $\mathfrak{iso}(1,3)$. The latter is composed of ten generators, the four generators of local translations $P_a, a=1,\ldots,4,$ and the six Lorentz transformations $M_{ab}$, satisfying the following commutation relations\footnote{We employ the standard convention that antisymmetrizations are taken with weight 1.}
\begin{equation}
[M_{ab},M_{cd}]=4\eta_{[a[c}M_{d]b]}~,\,\,\,\,\,
[P_a,M_{bc}]=2\eta_{a[b}P_{c]}~,\,\,\,\,\,
[P_a,P_b]=0~,
\end{equation}
where $\eta_{ab}$ is the (mostly plus) Minkowski metric. Gauging proceeds with the introduction of a gauge field for each algebra generator,
in particular the vielbein $e_{\mu}{}^a$ for translations and the spin connection $\omega_{\mu}{}^{ab}$ for Lorentz transformations.
The gauge connection is collectively given as
\begin{equation}
A_{\mu}(x)=e_{\mu}{}^a(x)P_a+\frac{1}{2}\omega_{\mu}{}^{ab}(x)M_{ab}~,\label{connection}
\end{equation}
and it transforms in the adjoint representation according to the standard rule
\begin{equation}
\delta A_{\mu}=\partial_{\mu}\epsilon+[A_{\mu},\epsilon]~,\label{transformconnection}
\end{equation}
where the gauge transformation parameter is taken to be
\begin{equation}
\epsilon(x)=\xi^a(x) P_a+\frac{1}{2} \lambda^{ab}(x)M_{ab}~.\label{parameter}
\end{equation}
Thus, replacing \eqref{connection} and \eqref{parameter} in \eqref{transformconnection}, one  finds the transformations of the vielbein and spin connection,
\begin{align}
\delta e_{\mu}{}^a&=\partial_{\mu}\xi^a+\omega_{\mu}{}^{ab}\xi_b-\lambda^a{}_be_{\mu}{}^b~,
\\
\delta \omega_{\mu}{}^{ab}&=\partial_{\mu}\lambda^{ab}-2\lambda^{[a}{}_c\omega_{\mu}{}^{cb]}~.
\end{align}
Their curvatures are obtained accordingly, using the usual formula
\begin{equation}
R_{\mu\nu}(A)=2\partial_{[\mu}A_{\nu]}+[A_{\mu},A_{\nu}]~.\label{usualformula}
\end{equation}
Writing $R_{\mu\nu}(A)=R_{\mu\nu}{}^a(e)P_a+\frac{1}{2} R_{\mu\nu}{}^{ab}(\omega)M_{ab}$ and replacing it along with \eqref{connection} in \eqref{usualformula}, the component tensors of curvature turn out to be
\begin{align}
R_{\mu\nu}{}^a(e)&=2\partial_{[\mu}e_{\nu]}{}^a-2\omega_{[\mu}{}^{ab}e_{\nu]b}~,\\
R_{\mu\nu}{}^{ab}(\omega)&=2\partial_{[\mu}\omega_{\nu]}{}^{ab}-2\omega_{[\mu}{}^{ac}\omega_{\nu]c}{}^b~.
\end{align}
Imposing the conventional curvature constraint $R_{\mu\nu}{}^a(e)=0$ (vanishing torsion), leads to the solution of the spin
connection in terms of the vielbein components.
These are certainly well-known and further details may be found in
textbooks. The dynamic follows from the Einstein-Hilbert action
\begin{equation}
S_{\mathrm{EH4}}=\frac{1}{2} \int d^4x\,\epsilon^{\mu\nu\rho\sigma}\epsilon_{abcd}\, e_{\mu}{}^{a}e_{\nu}{}^b R_{\rho\sigma}{}^{cd}(\omega)~.
\end{equation}
The latter action does not follow from gauge theory and this is the reason why the four-dimensional gravity cannot be considered as a gauge theory, even though the kinematics are correctly obtained in this gauge-theoretical approach.   

\subsection{Three-Dimensional case}

In the three-dimensional case, the first order formulation of general relativity can be recovered as a gauge theory of the Poincar\'e algebra $\mathfrak{iso}(1,2)$. If cosmological constant is present, the relevant algebras are the de Sitter and Anti de Sitter ones, $\mathfrak{so}(1,3)$ and $\mathfrak{so}(2,2)$,
respectively. The corresponding generators are the ones of local translations $P_a, a=1,2,3$ and the Lorentz transformations $J_{ab}$, satisfying the commutation relations \footnote{Again, we employ the standard convention that antisymmetrizations are taken with weight 1.}
\begin{equation}
[J_{ab},J_{cd}]=4\eta_{[a[c}J_{d]b]}~,\quad
[P_a,J_{bc}]=2\eta_{a[b}P_{c]}~,\quad
[P_a,P_b]=\Lambda J_{ab}~,
\end{equation}
where $\eta_{ab}$ is the (mostly plus) Minkowski metric and $\Lambda$ is the cosmological constant. These commutation relations are valid in any dimension\footnote{They are the same used in the four-dimensional case, with $\Lambda=0$.}, however, especially in three dimensions, a convenient rewriting is allowed, converting the above commutation relations to
\begin{equation}
[J_{ a},J_b]= \epsilon_{ a b c}J^{ c}~,\quad
[P_{ a},J_b]= \epsilon_{ a b c}P^{ c}~,\quad
[P_{ a},P_b]=\Lambda \epsilon_{abc}J^c~,
\end{equation}
using the definition $J^a=\frac{1}{2} \epsilon^{abc}J_{bc}$.
 Gauging proceeds with the introduction of a gauge field for each generator of the algebra, in particular the dreibein $e_{\mu}{}^a$ for translations and the spin connection $\omega_{\mu}{}^a=\frac{1}{2}\epsilon^{abc}\omega_{\mu bc}$ for Lorentz transformations.
The gauge connection is given as
\begin{equation}
A_{\mu}(x)=e_{\mu}{}^a(x)P_a+ \omega_{\mu}{}^{a}(x)J_{a}~,
\end{equation}
transforming in the adjoint representation according to the standard rule
\begin{equation}
\delta A_{\mu}=\partial_{\mu}\epsilon+[A_{\mu},\epsilon]~,
\end{equation}
where the gauge transformation parameter is 
\begin{equation}
\epsilon(x)=\xi^a(x) P_a+ \lambda^{a}(x)J_{a}~.
\end{equation}
Thus one can find the transformations of the dreibein and spin connection,
\begin{align}
\delta e_{\mu}{}^a&=\partial_{\mu}\xi^a-\epsilon^{abc}\left(\xi_b\omega_{\mu c}+\lambda_{b}e_{\mu c}\right)~,\label{deltae}
\\ 
\delta \omega_{\mu}{}^{a}&=\partial_{\mu}\lambda^{a}-\epsilon^{abc}\left(\lambda_b\omega_{\mu c}+\Lambda \xi_b e_{\mu c}\right)~,\label{deltao}
\end{align}
and their curvatures, using the standard formula
\begin{equation}
R_{\mu\nu}(A)=2\partial_{[\mu}A_{\nu]}+[A_{\mu},A_{\nu}]~.
\end{equation}
Writing $R_{\mu\nu}(A)=T_{\mu\nu}{}^aP_a+  R_{\mu\nu}{}^{a}J_{a}$\,, these turn out to be
\begin{align}\label{tmn}
T_{\mu\nu}{}^a&=2\partial_{[\mu}e_{\nu]}{}^a+2\epsilon^{abc}\omega_{[\mu b}e_{\nu]c}~,\\ \label{rmn}
R_{\mu\nu}{}^{a}&=2\partial_{[\mu}\omega_{\nu]}{}^{a}+\epsilon^{abc}\left(\omega_{\mu b}\omega_{\nu c}+\Lambda e_{\mu b}e_{\nu c}\right)~.
\end{align}
The Einstein-Hilbert action with or without cosmological constant, $\Lambda$, in three dimensions,
\begin{equation} \label{eh3}
S_{\mathrm{EH3}}=\frac{1}{16\pi G} \int_M  \epsilon^{\mu\nu\rho}\left(e_{ \mu}{}^{a}\left(\partial_{\nu}\omega_{\rho a}-\partial_{\rho}\omega_{\nu a}\right)+\epsilon_{a b c}e_{\mu}{}^a\omega_{\nu}{}^b\omega_{\rho}{}^c+\frac{1}{3} \Lambda \epsilon_{abc}e_{\mu}{}^ae_{\nu}{}^be_{\rho}{}^c
\right)\,,
\end{equation}
is identical to the action functional of a Chern-Simons gauge theory of the Poincar\'e, dS or AdS algebra, upon choice of an appropriate
quadratic form in the algebra \cite{AchucarroTownsend,WittenCS}. The standard choice is
\begin{equation}
\mathrm{tr}(J^aP^b)=\delta^{ab}~, \quad \mathrm{tr}(P^aP^b)=\mathrm{tr}(J^aJ^b)=0~.
\end{equation}
However, in three dimensions, for $\Lambda\neq 0$, there exists an alternative non-degenerate invariant quadratic form, given as
\begin{equation} \label{tr2}
\mathrm{tr}(J^aP^b)=0~, \quad \frac{1}{\Lambda}\mathrm{tr}(P^aP^b)=\mathrm{tr}(J^aJ^b)=\delta^{ab}~,
\end{equation}
yielding a different, yet classically equivalent, action \cite{WittenCS}. This second set of traces will be important in our study too.

Therefore, we conclude that in the three-dimensional case, gravity is successfully formulated as a gauge theory of Chern - Simons type, both for the transformations of the gauge fields and for its dynamics. 

\section{Noncommutative gauge theories}

In the noncommutative framework, gauge theories exist and are well-defined. Initially, one may consider an algebra, $\mathcal{A}$, of operators, $X_\mu$, being regarded as the noncommutative space and noncommutative coordinates, respectively. The noncommutative coordinates (operators) are subjected to a commutation relation which is generically given as
\begin{equation}
[X_\mu,X_\nu]=i\theta_{\mu\nu}\,.\label{noncomm}
\end{equation}
We should note that $\theta_{\mu \nu}$ determines the type of noncommutativity and is not a priori specified mainly for two reasons. First, although it is assumed to depend on the noncommutative coordinates $X_{\mu }$, it could in principle depend on the corresponding momenta $P_{\mu }$, as well (see Ref. \cite{Szabo:2009tn} for some examples). Second, one may consider noncommutative spaces where the coordinates $X_{\mu}$ do not close, as for example in the case of fuzzy 4-sphere of Ref. \cite{Castelino:1997rv}, or more generally spaces where $\theta_{\mu\nu}$ is not a fixed tensor, as for example in Refs. \cite{Yang:1947ud,Heckman:2014xha}. 


A very natural way to introduce gauge theories in the noncommutative regime is through the covariant noncommutative coordinates, $\mathcal{X}_\mu$\cite{Madore:2000en}, defined as
\begin{equation}
\mathcal{X}_\mu=X_\mu+A_\mu\,,
\end{equation}  
obeying a covariant gauge transformation rule 
\begin{equation}
\delta\mathcal{X}_\mu=i[\epsilon, \mathcal{X}_\mu]\,.\label{standardrule}
\end{equation}
The definition of the covariant noncommutative coordinate induces the definition of a noncommutative covariant field strength tensor which in turn defines the noncommutative gauge theory
\begin{equation}
F_{\mu\nu}=[\mathcal{X}_\mu,\mathcal{X}_\nu]-iC_{\mu\nu}^{~~~~\rho}\mathcal{X}_{\rho} \,,
\end{equation}
where the second term of the right hand side is specified for the Lie-type noncommutativity. In general, this term is determined by the type of noncommutativity of the coordinates $X_\mu$.   

Whether the gauge theory is Abelian or non-Abelian is determined by the nature of the gauge parameter, $\epsilon$. If $\epsilon$ is valued in the algebra $\mathcal{A}$, then the gauge theory is considered to be Abelian, while when $\epsilon$ is valued in $\text{Mat}(\mathcal{A})$, then the theory is non-Abelian \cite{Madore:2000en}. Given that we are going to deal with the non-Abelian case, it is a good point to settle a generic issue encountered in noncommutative non-Abelian gauge theories, that is to determine where the gauge fields are valued. Let us consider the following commutation relation 
\begin{equation}
[\epsilon,A]=[\epsilon^AT^A,A^BT^B]=\frac{1}{2} \{\epsilon^A,A^B\}[T^A,T^B]+\frac{1}{2} [\epsilon^A,A^B]\{T^A,T^B\}~, \label{nonAbelianCR}
\end{equation} 
where, $T^A$ are the generators of the algebra and the spacetime indices are suppresed for simplicity. In the commutative case, the last term of the above relation vanishes due to the commutator factor, which is trivially zero, since $\epsilon^A$ and $A^B$ are functions which depend on ordinary coordinates and therefore commute. That is the reason why there is no need to pay attention in the outcome of the anticommutator of the generators of the algebra. However, in the noncommutative case, the commutator of the last term is not zero, therefore one has to determine the nature of the outcome of the anticommutator. In general, the anticommutator of the generators, living in an arbitrary representation, does not give elements of the algebra but instead, products of the generators. Therefore, in general, restriction to a matrix algebra is not possible \cite{Madore:2000en} and further measures have to be applied in order to achieve the closure of the anticommutator. The first option is to consider the universal enveloping algebra \cite{Jurco:2000ja}, that is to consider an algebra that includes every product of generators produced by the anticommutator. The second option to overpass this drawback is to fix the representation so that the anticommutator will give a limited number of elements not belonging to the algebra and then  extend the algebra by those elements produced by the anticommutator in order to result with an enlarged -but finite- algebra in a fixed representation. In the noncommutative non-Abelian gauge theories we are going to build we will employ the second choice.

\section{Three-dimensional fuzzy spaces}

Based on the methodology we reviewed in the subsection 2.2 and the tools we briefly listed in section 3, our goal is to build a noncommutative non-Abelian gauge theory describing three-dimensional gravity. In order to do so, it is fundamental to determine the noncommutative spacetime we are going to work and build our gauge theory on.

A very well-known covariant noncommutative space is the fuzzy sphere \cite{Madore:1991bw, Hoppe}, which is defined in terms of three rescaled angular momentum operators $X_i=\lambda J_i$, the Lie algebra generators of a unitary irreducible representation of SU(2), which satisfy
\begin{equation}
[X_i,X_j]=i\lambda \epsilon_{ijk}X_k~,\quad \sum_{i=1}^3 X_iX_i=\lambda^2 j(j+1):=r^2~,\;\lambda \in \mathbb{R},\;2j\in \mathbb{N}.
\end{equation}   
If we choose to relax the Casimir condition, that is to allow the coordinates, $X_i$, to live in unitary reducible representations of SU(2) and keeping $\lambda$ fixed, the three-dimensional noncommutative space known as $\mathbb{R}^{3}_{\lambda}$ \cite{Hammou:2001cc} is obtained, expressed as a direct sum of fuzzy spheres with all possible radii determined by $2j\in\mathbb{N}$ \cite{Hammou:2001cc,Wallet:2016ilh,Vitale:2012dz,Vitale:2014hca} 
\begin{equation}
\mathbb{R}^3_\lambda=\sum_{2j\in\mathbb{N}} S^{2}_{\lambda, j}\,.
\end{equation}
Thus $\mathbb{R}^{3}_{\lambda}$ can be viewed as a discrete foliation of 3D Euclidean space by multiple fuzzy 2-spheres, each being a leaf of the foliation \cite{DeBellis:2010sy}. The structure and relation of $\mathbb{R}^{3}_{\lambda}$ to hermitian generators of $\mathfrak{su}(2)$ in a matrix basis appears e.g. in \cite{Wallet:2016ilh}.

The above construction of $\mathbb{R}_\lambda^3$ has a direct analogue in the case of Lorentzian signature. This means that three-dimensional Minkowski spacetime gets foliated by another two-dimensional fuzzy space, that is the fuzzy hyperboloid, $dS_2$. The latter is defined in a similar way to the fuzzy sphere, that is:
\begin{equation}
[X_i,X_j]=i\lambda C_{ij}{}^{k}X_k~,\quad \sum_{i,j}\eta_{ij}X_iX_j=\lambda^2 j(j-1)~,
\end{equation} 
where $X_i$ are matrices proportional to the generators of the SU(1,1) algebra $J_i$, $ C_{ij}{}^{k}$ are its structure constants and $\eta_{ij}$ is the three-dimensional Minkowskian metric. It is important to mention that for the construction of $dS_F^2$ the irreducible representations of the group are chosen appropriately, specifically from the continuous principle series \cite{Jurman:2013ota}. 

Again, in order to obtain the Minkowskian analogue of $\mathbb{R}_\lambda^3$, one has to lift the Casimir condition and consider the $X_i$ to live in reducible representations, with each block being an irreducible representation, i.e. a fuzzy hyperboloid of fixed radius. This means that three-dimensional Minkowski spacetime gets foliated by the fuzzy hyperboloids of various radii. 

It is worth-noting that although the two constructions are similar, there is a major difference between them. In the Euclidean case the whole construction is based on SU(2) and its unitary irreducible representations, which are finite-dimensional since SU(2) is compact. On the other hand, in the Minkowskian case, the structure is based on SU(1,1) which possesses only infinite-dimensional unitary irreducible representations, since it is non-compact. 
     
\section{Gravity as a gauge theory on the three-dimensional fuzzy spaces}

In section 2.2, we reviewed the description of three-dimensional gravity as a Chern-Simons gauge theory. In this section, we propose a description for a noncommutative version of three-dimensional gravity on the fuzzy spaces we described in section 4, employing the same methodology of the commutative case, using the tools of noncommutative gauge theories, mentioned in section 3. Therefore, according to the gauging procedure, the covariant coordinate should contain information about the noncommutative versions of the dreibein and the spin connection \footnote{For similar approaches see \cite{Nair:2001kr,Abe:2002in,Nair:2006qg} in the same sense that covariant derivative does in the gauging of Poincar\'e and (A)dS algebra in the commutative case.}.    

\subsection{The Lorentzian case}

Since we are going to deal with the three-dimensional case with $\Lambda > 0$, the relevant  group of symmetry of the fuzzy space we are using is SO(1,3) (SO(4) for the euclidean case of $\mathbb{R}_\lambda^3$). This consideration will lead to a non-Abelian noncommutative gauge theory, which means that we will have to overcome the drawback mentioned and explained in section 3, extending the algebra appropriately. Our approach is motivated by the one followed in \cite{Aschieri:2009ky} in the Moyal-Weyl case. Accordingly, first we consider the spin group of the group of symmetry, which is the $\text{Spin}(1,3)$ being isomorphic to the SL(2;$\mathbb{C}$) (similarly in the Euclidean case it is $\text{Spin}(4)=\text{SU}(2)\times \text{SU}(2)$). The generators of SL(2;$\mathbb{C}$), are $\Sigma_{AB}=\frac{1}{2}\gamma_{AB}=\frac{1}{4}[\gamma_A,\gamma_B], A=1,2,3,4$, $\gamma_A$ being 4D Lorentzian gamma matrices, satisfying the following commutation relation:
\begin{equation}
[\gamma_{AB},\gamma_{CD}] = 8\eta_{[A[C}\gamma_{D]B]}~,
\end{equation}
found through the product relation \cite{VanProeyen:1999ni}:
\begin{equation}
\gamma_{AB}\gamma^{CD}=2\delta^{[C}_{[B}\delta^{D]}_{A]}+4\delta^{[C}_{[B}\gamma_{A]}{}^{D]}+i\epsilon_{AB}{}^{CD}\gamma_5~.
\end{equation}
The generators of the algebra do not anticommute, yielding elements outside the algebra. Therefore, we fix the representation to the be the spinorial, which enables us to control the elements produced by the anticommutators to the minimum of two: the unit matrix, $\one$, and $\gamma_5$, due to the above product relation:
\begin{equation}
\{\gamma_{AB},\gamma_{CD}\} = 4\eta_{C[B}\eta_{A]D} {\one} +2i\epsilon_{ABCD}\gamma_5\,.
\end{equation}  
Therefore, the algebra has to be extended including these two elements, leading to an eight-dimensional algebra, which is in fact the $\mathrm{GL}(2;\mathbb{C})$, generated by\footnote{We use the same set of $\gamma_4$-hermitian generators as in Ref. \cite{Aschieri:2009ky} (according to our conventions, $\gamma_4$ corresponds to the $\gamma_0$ of that paper, and our metric signature is the opposite one, with $\eta_{44}=-1$); see also \cite{Aschieri:2012vf}, Sec. 4.2, for a detailed explanation.} $\{\gamma_{AB},\gamma_5,i\one\}$ \footnote{In the Euclidean case, a similar extension takes place, $\text{SU}(2)\times \text{SU}(2)$ symmetry gets enlarged to the $\text{U}(2)\times \text{U}(2)$.}.

We proceed with a decomposition to an $\text{SO}(3)$ notation, obtaining the generators $\gamma_{ab}$ and $\gamma_a=\gamma_{a4}$ with $a=1,2,3$. A notational simplification is achieved after defining: $\widetilde\gamma^a=\epsilon^{abc}\gamma_{bc}$. In order to proceed we need to conform the commutation and anticommutation relations to the new definitions of the generators after the SO(3) decomposition:
\begin{align}
&[\tilde{\gamma}^a,\tilde{\gamma}^b]=-4\epsilon^{abc}\tilde{\gamma}_c\,, \,
[\gamma_a,\tilde{\gamma}_b]=-4\epsilon_{abc}\gamma^c\,, \, [\gamma_a,\gamma_b]=\epsilon_{abc}\tilde{\gamma}^c\,,\\
&\{\tilde{\gamma}^a,\tilde{\gamma}^b\}=-8\eta^{ab}{\one}\,,\, \{\gamma_a,\tilde{\gamma}^b\}=4i\delta_a^b\gamma_5\,,\, \{\gamma_a,\gamma_b\}=2\eta_{ab}{\one}\,, \\
&[\gamma^5,\gamma^{AB}]=0\,,\, \{\gamma^5,\gamma^{AB}\}=i\epsilon^{ABCD}\gamma_{CD}~,\, \{\gamma_a,\gamma_5\}=i\tilde{\gamma}_a~,\, \{\tilde{\gamma}_a,\gamma_5\}=-4i\gamma_a~.
\end{align}

We resume our analysis by considering $\text{GL}(2;\mathbb{C})$ as the gauge group. The noncommutative coordinates $X_{a}$ are identified with the three operators of the 3D fuzzy space discussed in Section 4. Therefore, following the discussion of Section 3 about noncommutative gauge theories, the covariant coordinates will incorporate the information of the deformation of the space, via the gauge connection, $\mathcal{A}_\mu$:
\begin{equation}
{\cal X}_{\mu}=\delta_{\mu}{}^aX_a+{\cal A}_{\mu}~,
\end{equation}
where ${\cal A}_{\mu}={\cal A}_{\mu}^{\bar{a}}(X_a)\otimes T^{\bar{a}}$, ${\cal A}_{\mu}^{\bar{a}}$ being the $\text{GL}(2;\mathbb{C})$-valued gauge fields. The tensor product is present because component fields are no longer functions of coordinates of classical manifold, but operators. The $\text{GL}(2;\mathbb{C})$ gauge connection is written in terms of the component gauge fields as follows:
\begin{equation}
{\cal A}_{\mu}(X)=e_{\mu}{}^a(X)\otimes \gamma_a+ \omega_{\mu}{}^{a}(X)\otimes \widetilde\gamma_{a}+{A}_{\mu}(X)\otimes i\one+\widetilde{A}_{\mu}(X)\otimes \gamma_5~.
\end{equation}
Accordingly, the gauge parameter is element of the algebra, therefore it is expanded on its generators, that is
\begin{equation}
\epsilon(X)=\xi^a(X)\otimes \gamma_a+ \lambda^{a}(X)\otimes \widetilde\gamma_{a}+\epsilon_0(X)\otimes i\one+\widetilde\epsilon_0(X)\otimes\gamma_5~.
\end{equation}

Using the general form of the covariant transformation rule\footnote{More precisely, due to the choice of the conventions for the GL(2,$\mathbb{C}$) generators, including $i\one$, we use here $\delta{\cal X}=[\epsilon,{\cal X}]$, dropping an $i$ from the standard rule \eqref{standardrule}.}, we calculate the transformations of the component gauge fields, in a similar way to the commutative case. The transformations are calculated to be (denoting $X_{\mu}=\delta_{\mu}{}^aX_a$):
\begin{align}
\delta e_\mu^{~a}&= -i[X_{\mu}+A_\mu,\xi^a]-2\{\xi_b,\omega_{\mu c}\}\epsilon^{abc}
						-2\{\lambda_b,e_{\mu c}\}\epsilon^{abc}+i[\epsilon_0,e_\mu^{~a}]-\nonumber\\ 
						&-2i[\lambda^a,\widetilde{A}_{\mu}]-2i[\widetilde{\epsilon}_0,\omega_{\mu}{}^a]~,\\
\delta\omega_\mu^{~a}&= -i[X_{\mu}+ A_\mu,\lambda^a]
							+\frac{1}{2}\{\xi_b,e_{\mu c}\}\epsilon^{abc}-2\{\lambda_b,\omega_{\mu c}\}\epsilon^{abc}+i[\epsilon_0,\omega_\mu^{~a}]+\nonumber\\ 
							&+\frac{i}{2} [\xi^a,\widetilde{A}_{\mu}]+\frac{i}{2}[\widetilde{\epsilon}_0,e_{\mu}{}^a]~,\\
\delta {A}_\mu&=-i[X_{\mu}+{A}_\mu,\epsilon_0]
						-i[\xi_a,e_\mu^{~a}]+4i[\lambda_a,\omega_\mu^{~a}]
						-i[\tilde{\epsilon}_0,\widetilde{A}_\mu]~,\label{deltaA}\\
\delta\widetilde{A}_\mu&=-i[X_{\mu}+{A}_\mu,\tilde{\epsilon}_0]
						+ 2i[\xi_a,\omega_\mu^{~a}]+2i[\lambda_a,e_\mu^{~a}]+i[\epsilon_0,\widetilde{A}_\mu]~,
\end{align}
where we have been cautious about the order of the generators (unlike the commutative case) and have used the formula \eqref{nonAbelianCR}. 

Let us consider two limits for the above transformation rules. First, the Abelian limit, that is the case in which we would have considered an Abelian gauge group. In this case we would have obtained just an Abelian gauge theory on the 3D fuzzy space. This effectively amounts to setting $e_{\mu}{}^a=\omega_{\mu}{}^a=0$ and $\widetilde{A}_{\mu}=0$, having only one non-vanishing gauge parameter, that is the $\epsilon_0$, therefore only Eq. \eqref{deltaA} is non-trivial and it becomes:
\begin{equation}
\delta A_{\mu}=-i[X_{\mu},\epsilon_0]+i[\epsilon_0,A_{\mu}]~,
\end{equation}
which is the anticipated transformation rule of a noncommutative Maxwell gauge field. Thus we observe that the Maxwell sector is present, not depending on the triviality of the dreibein, with $X_{\mu}+{A}_\mu$ being the covariant coordinate.

Second, the commutative limit, in which the Yang-Mills and gravity fields disentangle, meaning that the fields introduced due to noncommutativity, $A_\mu, \tilde{A}_\mu$, vanish in this limit. Taking into consideration that $A_\mu$ vanishes and that the inner derivation becomes $[X_\mu, f]\to -i\partial_\mu f$, we obtain the following transformations for the dreibein and spin connection:
\begin{align}
\delta e_\mu^{~a}&=-\partial_\mu\xi^a-4\xi_b\omega_{\mu c}\epsilon^{abc}-4\lambda_be_{\mu c}\epsilon^{abc}~,\\
\delta\omega_\mu^{~a}&=-\partial_\mu\lambda^a +\xi_be_{\mu c}\epsilon^{abc}-4\lambda_b\omega_{\mu c}\epsilon^{abc}~.\label{cl}
\end{align}
After the redefinition of the fields, generators and parameters, that is $\gamma_a\rightarrow \frac{2i}{\sqrt{\Lambda}} P_a~, \tilde\gamma_a\to -4J_a~,$ and also
 $4\lambda^a\rightarrow \lambda^a,\;\xi^a\frac{2i}{\sqrt\Lambda}\rightarrow -\xi^a,\;e^a_\mu\rightarrow\frac{\sqrt\Lambda}{2i}e^a_\mu, \;\omega^a_\mu\rightarrow -\frac{1}{4}\omega^a_\mu$, the  transformation rules coincide to Eqs. \eqref{deltae} and \eqref{deltao}. Therefore, we confirm that in the commutative limit, the transformations of \cite{WittenCS} of the gauge fields of three-dimensional gravity are recovered.

The next step is to calculate the commutator of the covariant coordinates in order to obtain the curvature tensors. Since we are dealing with a case in which the right-hand side of the commutator of the coordinates is linear in generators, an additional linear term is included in the definition of curvature, that is:
\begin{equation}
\mathcal{R}_{\mu\nu}(X)=[\mathcal{X}_\mu,\mathcal{X}_\nu]-i\lambda C_{\mu\nu}{}^{\rho}\mathcal{X}_\rho~. 
\end{equation}
The curvature tensor is valued in the algebra of GL(2,$\mathbb{C}$), therefore it can be expanded on its generators as:
\begin{equation}
\mathcal{R}_{\mu\nu}(X)=T^a_{\mu\nu}(X)\otimes \gamma_a+R_{\mu\nu}^a(X) \otimes \tilde{\gamma}_a+F_{\mu\nu}(X)\otimes i\one+\widetilde{F}_{\mu\nu}(X)\otimes\gamma_5\,.
\end{equation}
Therefore, the various tensors are calculated as:
\begin{align}
T^a_{\mu\nu}&=i[X_{\mu}+{A}_\mu,e_\nu^{~a}]-i[X_{\nu}+{A}_\nu,e_\mu^{~a}]-2\{e_{\mu b},
				\omega_{\nu c}\}\epsilon^{abc}-2\{\omega_{\mu b},e_{\nu c}\}\epsilon^{abc}-\nonumber\\ 
				&\quad -2i[\omega_{\mu}{}^a,\widetilde{A}_{\nu}]+2i[\omega_{\nu}{}^a,\widetilde{A}_{\mu}]-i\lambda C_{\mu\nu}^{~~~\rho}e^{~a}_\rho~,
\\
R^a_{\mu\nu}&=i[X_{\mu}+{A}_\mu,\omega_\nu^{~a}]-i[X_{\nu}+{A}_\nu,\omega_\mu^{~a}]
			-2\{\omega_{\mu b},\omega_{\nu c}\}\epsilon^{abc}+\frac{1}{2}\{e_{\mu b},e_{\nu c}\}\epsilon^{abc}+\nonumber\\ & +\frac{i}{2}[e_{\mu}{}^a,\widetilde{A}_{\nu}]-\frac{i}{2}[e_{\nu}{}^a,\widetilde{A}_{\mu}]-i\lambda C_{\mu\nu}^{~~~\rho}\omega_\rho^{~a}~,\\
F_{\mu\nu}&=i[X_{\mu}+{A}_\mu,X_{\nu}+{A}_\nu]-i[e_\mu^{~a},e_{\nu a}]+
			4i[\omega_\mu^{~a},\omega_{\nu  a}]-i[\widetilde{A}_\mu,\widetilde{A}_\nu]-i\lambda C_{\mu\nu}^{~~~\rho}(X_{\rho}+{A}_\rho)~,\\
\widetilde{F}_{\mu\nu}&=i[X_{\mu}+{A}_\mu,\widetilde{A}_\nu]-i[X_{\nu}+{A}_\nu,\widetilde{A}_\mu]+2i[e_\mu^{~a},\omega_{\nu a}]+2i[\omega_\mu^{~a},e_{\nu a}]-i\lambda C_{\mu\nu}^{~~~\rho}\widetilde{A}_\rho~.\label{tensors}
\end{align}
Once more, in the commutative limit the expected results appearing in Eqs.
\eqref{tmn} and \eqref{rmn} are recovered, after applying the aforementioned rescalings.

\subsection{The Euclidean case} 

As explained in the previous subsection, in the Euclidean case one has to consider the $\text{U}(2)\times \text{U}(2)$ to be the gauge group with its representation fixed. Recalling that U(2) is spanned by four generators, i.e. the Pauli matrices and the unit matrix, one understands that the expansions of the gauge field and the gauge parameter should involve the following 4$\times$4 matrices:
\begin{align}
J^L_a=\begin{pmatrix}
	\sigma_a & 0 \\ 0 & 0
\end{pmatrix}~, \quad J^R_a=\begin{pmatrix}
0 & 0 \\ 0 & \sigma_a
\end{pmatrix}~,
\end{align}
and
\begin{align}
J^L_0=\begin{pmatrix}
	\one & 0 \\ 0 & 0
\end{pmatrix}~, \quad J^R_0=\begin{pmatrix}
	0 & 0 \\ 0 & \one
\end{pmatrix}~.
\end{align}
However, one should be careful with the identification of the noncommutative dreibein and spin connection in the expansion of the gauge field. The correct interpretation is achieved after considering the following linear combination of the above matrices:
\begin{align}
P_a=\frac{1}{2} (J_a^L-J_a^R)=\frac{1}{2} \left(\begin{array}{cc}
	\sigma_a & 0 \\ 0 & -\sigma_a
\end{array}\right)~, \quad M_a=\frac{1}{2}(J_a^L+J_a^R)=\frac 12 \left(\begin{array}{cc}
\sigma_a & 0 \\ 0 & \sigma_a
\end{array}\right)~,
\end{align}
and also
\begin{equation}
\one=J_0^L+J_0^R~, \quad \gamma_5=J_0^L-J_0^R~.
\end{equation}
These are the generators that satisfy the expected commutation and anticommutation relations:
\begin{align}
&[P_a,P_b]=i\epsilon_{abc}M_c~, \quad [P_a,M_b]=i\epsilon_{abc}P_c~, \quad [M_a,M_b]=i\epsilon_{abc}M_c~,
\\
& \{P_a,P_b\}=\frac{1}{2} \delta_{ab}\one~,\quad \{P_a,M_b\}=\frac{1}{2} \delta_{ab}\gamma_5~, \quad \{M_a,M_b\}=\frac{1}{2} \delta_{ab}\one~.
\\
& [\gamma_5 , P_a] = [\gamma_5, M_a] = 0~,\quad
\{\gamma_5 , P_a\} =2 M_a~,\quad  \{\gamma_5, M_a\} = 2 P_a~.
\end{align}
One then proceeds in a similar way to the Lorentzian case, with the covariant coordinate being:
\begin{equation}
{\cal X}_{\mu}(X)= X_{\mu}\otimes i\one +e_{\mu}{}â(X)\otimes P_a+\omega_{\mu}{}â(X)\otimes M_a+A_{\mu}(X)\otimes i\one+\widetilde{A}_{\mu}(X)\otimes \gamma_5~~,
\end{equation}
and the gauge parameter given as:
\begin{equation}
\epsilon(X)=\xi^a(X)\otimes P_a+\lambda^a(X)\otimes M_a+\epsilon_0(X)\otimes i\one+\widetilde{\epsilon}_0(X)\otimes\gamma_5~.
\end{equation}
The only difference to the Lorentzian case is the metric signature, thus it is redundant to rewrite the results.

\subsection{Action of 3D fuzzy gravity}

Finally, we give an action of the present noncommutative three-dimensional gravity model. In Ref. \cite{Jurman:2013ota}, it is shown that fuzzy 2-hyperboloids give dynamical brane solutions of a Yang-Mills type matrix model with the characteristic term of squared commutator. From our point of view, since we work in three dimensions and we know that general relativity has no dynamics, we propose the following action (cf. \cite{Gere:2013uaa})\footnote{A similar action was proposed in Ref. \cite{Valtancoli:2003ve} for a gravity theory on the fuzzy sphere. See also \cite{Alekseev:2000fd}.}:
\begin{equation} 
S_0= \frac{1}{g^2}\text{Tr}\left(\frac{i}{3}C^{\mu\nu\rho}{X}_\mu{ X}_\nu{X}_\rho
-m^2{ X}_{\mu}{ X}^{\mu} \right)~. 
\end{equation}
The three-dimensional fuzzy space we considered is indeed a solution of the field equations derived from the above action,
\begin{equation} 
[X_{\mu},X_{\nu}]-2im^2C_{\mu\nu}{}^{\rho}X_{\rho}=0~,
\end{equation}
 when $2m^2=\lambda$.

Furthermore, we would like to write the action including the gauge fields. One could either consider the fluctuations around the above solution, or directly write down an action for the curvatures in the spirit of \cite{Madore:2000en}.
This action should be written in terms of the covariant coordinates ${\cal X}_{\mu}$ and it should also contain a prescription for taking the trace over the gauge algebra. Regarding this matter, although there are two different trace prescriptions available \cite{WittenCS}, only one of them works in our case. This is because we have fixed the representation and used gamma matrices in our expansions. Thus the prescription imposed on us by the algebra of gamma matrices is the one corresponding to Eq.\eqref{tr2}.
More specifically, we use the trace relations
\begin{equation}
\mathrm{tr}\left(\gamma_a\gamma_b\right)= 4\eta_{ab}~,\quad \text{tr}\left(\tilde\gamma_a\tilde\gamma_b\right)= -16\eta_{ab}~. 
\end{equation}
Therefore, the action we propose is
\begin{equation} 
S=\frac{1}{g^2}\text{Tr}~\text{tr}\left(\frac{i}{3}C^{\mu\nu\rho}{\cal X}_\mu{\cal X}_\nu{\cal X}_\rho
-\frac{\lambda}{2}{\cal X}_{\mu}{\cal X}^{\mu} \right)~, \label{action!}
\end{equation}
where the first trace $\text{Tr}$ is over the matrices $X$ and the second trace $\text{tr}$ is over the algebra.
We can rewrite this action as
\begin{equation}
S=\frac{1}{6g^2} \text{Tr}~\text{tr}\left(iC^{\mu\nu\rho}{\cal X}_\mu {\cal R}_{\nu\rho}\right)+S_{\lambda}~,\label{action2}
\end{equation}
where $S_{\lambda}=-\frac{\lambda}{6g^2}\text{Tr}~\text{tr}\left({\cal X}^{\mu}{\cal X}_{\mu}\right)$ and it vanishes in the limit $\lambda \to 0$.
Using the explicit form of the algebra trace, the first term in the action is proportional to
\begin{equation}
\text{Tr}~C^{\mu\nu\rho}(e_{\mu a}T^a_{\nu\rho}-4\omega_{\mu a}R^a_{\nu\rho}-(X_{\mu}+{ A}_\mu) F_{\nu\rho}+\widetilde A_\mu\widetilde F_{\nu\rho})~.
\end{equation}
This action is similar to the one obtained in Section 2.3 of Ref. \cite{WittenCS}. Upon taking the commutative limit and performing again the redefinitions, the first two terms are identical to that action; however, in the present case we necessarily obtain an additional sector, associated to the additional gauge fields that cannot decouple in the noncommutative case. 

Variation of the above action, \eqref{action!}, with respect to the covariant coordinate gives the equations of motion
\begin{equation}
T_{\mu\nu}^{~~~a}=0\,,\quad\quad R_{\mu\nu}^{~~~a}=0\,,\quad\quad F_{\mu\nu}=0\,,\quad\quad \tilde{F}_{\mu\nu}=0
\end{equation}
It is worth-noting that the same equations of motion are obtained after variation with respect to the gauge fields of the action in the form \eqref{action2}, after using the algebra trace and replacing with the explicit expressions of the component tensors \eqref{tensors}.

\section{Conclusions}

In this review we presented a model for three-dimensional gravity on two specific fuzzy spaces, the $\mathbb{R}_\lambda^3$ and its Lorentzian analogue. In order to make it self-contained, we included the corresponding commutative well-established model of three-dimensional gravity obtained as a gauge theory of the isometry group of the three dimensional Minkowski or (A)dS spacetime. Combining this formulation with the existence of gauge theories on noncommutative spaces, via the definition of the covariant coordinate, we obtained a Chern-Simons type action for three-dimensional noncommutative gravity and the corresponding equations of motion. 

More specifically the two three-dimensional noncommutative spaces we considered are the $\mathbb{R}_\lambda^3$ and its Lorentzian analogue. The first one is the discrete foliation of the Euclidean space by fuzzy spheres and is based on reducible representations of the SU(2) having a natural SO(4) symmetry, while the latter is the discrete foliation of Minkowski spacetime by fuzzy hyperboloids based on reducible representations of the SU(1,1) group having an SO(1,3) symmetry (for positive cosmological constant).

Then, we considered gauge theories on the above fuzzy spaces, with gauge groups the ones of their underlying symmetries. The fact that the symmetry groups are non-Abelian resulted into the consideration of fixed representations as well as the extension of the two algebras in order that the anticommutators close, that is to yield elements within the algebra. More specifically the SO(1,3) was extended to the GL(2,$\mathbb{C}$), while the SO(4) to the U(2)$\times$U(2). We should note that noncommutative gauge theories describing gravity have been considered before with gauge group the GL(2,$\mathbb{C}$), but for a four-dimensional case, on the Moyal-Weyl space \cite{Chamseddine:2003we,Aschieri:2009ky}.    

Then, following the standard procedure of building gauge theories on noncommutative spaces, we defined the covariant coordinate and along with the consideration of a Lie-valued parameter, we ended up with the transformations of the gauge fields. The definition of the covariant coordinate led to the calculation of the corresponding curvatures and finally to a matrix action of Chern-Simons type, which after variation gave the equations of motion. It is worth-noting that the above results reproduce the expressions of the commutative limit, after taking the corresponding limit.   

The present work is the first step of a long-plan project which lies in four directions. The first one is a further analysis of the Lorentzian analogue of the $\mathbb{R}_\lambda^3$, that is to obtain the algebra of its functions and its differential structure. The second would be an effort to write the action in a specific matrix basis and study its behaviour using perturbation theory \cite{Lizzi:2014pwa}. The third direction is the upgrade of the present noncommutative three-dimensional gravity model to a four-dimensional one, modifying the procedure of the first. The general idea is to work on an extended $\mathbb{R}_\lambda^3$ four-dimensional space and apply the gauge theory procedure on it. The last direction is to incorporate the gauge group and the four-dimensional non-commutative space into a larger symmetry group, in the spirit of  \cite{Heckman:2014xha, Buric:2015wta}.

\paragraph{Acknowledgements}
We would like to acknowledge A.~Chatzistavrakidis and L.~Jonke for our fruitful collaboration and their contribution in the paper on which the present proceeding is based. Also, we would like to thank P.~Vitale for useful discussions. The work of D.J. was supported by the Croatian Science Foundation under the project IP-2014-09-3258 and by the H2020 Twinning Project No. 692194 "RBI-T-WINNING". G.Z. thanks the MPI Munich and LMU for hospitality and the A.v.Humboldt Foundation for support. G.M. thanks the Division of Theoretical Physics of Rudjer Bo$\check{\text{s}}$kovi\'c Institute in Zagreb, Croatia for the hospitality during a visit. Also, the same for a visit at the Physics Dept. of University Federico II in Napoli, Italy. The two visits of G.M. were supported by the COST action QSPACE MP1405.

\end{document}